\documentclass[twocolumn,superscriptaddress,showpacs,floatfix]{revtex4}

 \usepackage{graphicx}

\usepackage{bm}

\begin{document}

\title{Understanding the different scaling behavior in various shell models proposed for
turbulent thermal convection}
\author{Emily S.C. Ching}
\email{ching@phy.cuhk.edu.hk}
\affiliation{Department of Physics, The Chinese University of Hong Kong, Shatin,
New Territories, Hong Kong}
\affiliation{The Institute of Theoretical Physics, The Chinese University of Hong Kong, Shatin,
New Territories, Hong Kong}
\author{H. Guo }
\author{W.C. Cheng}
\affiliation{Department of Physics,
The Chinese University of Hong Kong, Shatin, New Territories, Hong
Kong}

\begin{abstract}
Different scaling behavior has been reported in various shell models proposed for
turbulent thermal convection. In this paper, we 
show that buoyancy is not always relevant to the statistical properties 
of these shell models even though there is an explicit coupling between velocity and
temperature in the equations of motion. When buoyancy is relevant (irrelevant) to the statistical
properties, the scaling behavior is Bolgiano-Obukhov (Kolmogorov) plus intermittency
corrections. We show that the intermittency corrections of temperature could be 
solely attributed to fluctuations in the entropy transfer rate 
when buoyancy is relevant but due to
fluctuations in both energy and entropy transfer rates when buoyancy is irrelevant. 
This difference can be used as a criterion to distinguish whether temperature is 
behaving as an active or a passive scalar.
\end{abstract}

\pacs{47.27-i,47.27te}


\maketitle


\section{\label{intro}Introduction}


Turbulent thermal convection is a problem of great research
interest~(see, for example,~\cite{Siggia,Kadanoff} for a review).
One interesting issue is to understand the scaling behavior of the
velocity and temperature fluctuations. Turbulent thermal
convection is often investigated experimentally in
Rayleigh-B{\'e}nard convection cells, which are closed cells of
fluid heated from below and cooled on the top. Such confined
turbulent convective flows are highly inhomogeneous as thermal and
viscous boundary layers are formed near the top and bottom of the
cell. Scaling laws for the central region of such confined
turbulent thermal convection have been put forth and shown to be in
good agreement with the existing experimental
measurements~\cite{Ching2007}. On the other hand, shell models
focussing on the energy cascade process have been studied
intensively and proved to be useful for understanding the scaling
behavior of velocity fluctuations in inertia-driven
turbulence~(see, for example,~\cite{Luca} for a review). It is
thus natural to also construct shell models for turbulent thermal
convection. Shell models are, by construction, boundary-free and
thus shell models for turbulent thermal convection are necessarily
models of {\it homogeneous} turbulent thermal convection. It is
known that the presence of boundaries generates coherent
structures such as plumes and a  large-scale mean flow in confined
turbulent thermal convection, and these coherent structures can
affect the scaling behavior~\cite{Ching2007}. Thus, scaling
behavior in confined turbulent thermal convection and scaling
behavior in homogeneous turbulent thermal convection as studied in
shell models can be different.

Several shell models for turbulent thermal convection have been
proposed and different scaling behavior reported.  Specifically,
Bolgiano-Obukhov (BO) scaling~\cite{BO} plus intermittency
corrections has been reported in the shell model constructed by
Brandenburg~\cite{Brandenburg} and also in the modified model by
Suzuki and Toh~\cite{Suzuki} for some parameter range. 
On the other hand, Kolmogorov 1941
(K41) scaling~\cite{K41} plus intermittency corrections has been
reported by Jiang and Liu~\cite{Liu} using a shell model extended
from the Gledzer-Ohkitani-Yamada (GOY) model~\cite{GOY}, which we
shall denote as the GOYT model. In this paper, we 
show that buoyancy is not always significant and directly relevant to
the statistical properties even though there is an explicit 
coupling between velocity and temperature in the equations of motion in all these
shell models. We clarify that the two different types of scaling
behavior reported correspond respectively to the case when
buoyancy is relevant to the statistical properties and the case when it is not. 
Specifically, the scaling behavior is BO plus intermittency corrections when buoyancy is
relevant, and K41 plus intermittency corrections
(as one would expect for temperature behaving as a passive scalar) 
when buoyancy is irrelevant.
We show that the intermittency corrections of temperature could be
solely attributed to fluctuations in the entropy transfer rate
when buoyancy is relevant but due to fluctuations in both energy and 
entropy transfer rates when buoyancy is irrelevant. This difference might
be used as a criterion to distinguish whether temperature is behaving as an active or
a passive scalar.

\section{\label{model}Shell models proposed for turbulent thermal convection}

Two classes of shell models have been proposed for studying
turbulent thermal convection. The first class consists of the
shell model proposed by Brandenburg~\cite{Brandenburg} and its
modified versions~\cite{Suzuki}. The other class consists of the
GOYT model, the shell model extended from the GOY model~\cite{Liu}
and the SabraT model~\cite{GuoThesis} from the Sabra
model~\cite{GuoThesis}. The Sabra model~\cite{Sabra} was proposed
to eliminate some undesirable periodic oscillations in the GOY
model, and have essentially the same scaling behavior as the GOY model. 
The scaling behavior in the first class of shell models is BO plus corrections
in some parameter range while the scaling behavior in the second class of shell models
is always K41 plus corrections.  In this paper, we focus on 
two shell models, one from each class. The first
one, denoted as the Brandenburg model, is the modified
model proposed by Suzuki and Toh~\cite{Suzuki} without the drag term. 
The second is the SabraT model.

The basic idea of a shell model is to consider variables in discrete
``shells"  in Fourier $k$-space, and construct a set of ordinary
differential equations for these variables per shell. For shell
models for turbulent thermal convection, there are two variables,
the velocity and temperature variables, $u_n$ and $\theta_n$. They
can be roughly thought of as the Fourier transforms of the
velocity and temperature fields with wavevector $\vec{k}$, whose
magnitude satisfies $k_n \le |\vec{k}| \le k_{n+1}$. Here, $k_n =
2^{n}k_0$ is the wavenumber of the $n$th shell, with $0 \le n \le N-1$,
and $k_0=1$ is the wavenumber corresponding to the largest scale
in the system. The equations of motion for $u_n$ and $\theta_n$
are:
\begin{eqnarray}
{{du_n }  \over {dt}} &=& I_u(k_n) -\nu k^2_n u_n +\alpha g \theta_n
\label{Velocity} \\
{{d\theta_n } \over {dt}} &=& I_\theta(k_n)  -\kappa k^2 _n
\theta_n + f_n \label{Temperature}
\end{eqnarray}
where $f_n$ is the forcing term acting only on the first few
shells. The nonlinear terms $I_u(k_n)$ and $I_\theta(k_n)$ are
taken to couple quadratically with  the nearest shells and
sometimes also the next nearest shells, and are constructed to
satisfy two conservation laws of energy and entropy
(proportional to $|\theta_n|^2$) in the limit of $\nu \to 0$ and
$\kappa \to 0$:
\begin{eqnarray}
{d \over {dt}} \left[{1 \over 2}\sum\limits_{n = 1}^N {|u_n|^2 }\right]  -
\alpha g\sum\limits_{n = 1}^N {Re\{u_n \theta^* _n} \}  &=& 0  \\
{d \over {dt}}\left[{1 \over 2}\sum\limits_{n = 1}^N {|\theta _n|^2 }\right]
&=& 0 
\end{eqnarray}
As a result, the nonlinear terms $u_n^*I_u(k_n)$ and
$\theta_n^*I_\theta(k_n)$ should have a fluxlike form such that
the evolution equations of energy and entropy in the $n$th shell are:
\begin{eqnarray}
\label{Eu}{d \over {dt}}\left[\frac{|u_n|^2}{2}\right] 
&=&F_u(k_n)-F_u(k_{n+1})- \nu k^2 _n |u_n|^2 \\ 
\nonumber && +\alpha g Re\{u_n \theta^*_n\} \\
\label{ET}
{d \over {dt}}\left[\frac{|\theta_n|^2}{2}\right]&=&F_\theta(k_n)-F_\theta(k_{n+1})- \kappa k^2 _n
|\theta_n|^2 +f_n\theta_n^*  \ \ \ 
\end{eqnarray}
The fluxes $F_u(k_n)$ and $F_\theta(k_n)$ are respectively the rates
of energy and entropy transfer from the $n-1$th shell to the $n$th shell.

In the Brandenburg model, $u_n$ and $\theta_n$ are real variables
with~\cite{Brandenburg,Suzuki}:
\begin{eqnarray}
\nonumber
I^B_u(k_n)&=& ak_n ( u^2 _{n - 1}  - 2u_n
u_{n + 1} ) \\ && + bk_n (u_n u_{n - 1}  - 2u^2 _{n + 1} )\\
I^B_\theta(k_n)&=&  \tilde ak_n ( u_{ n - 1} \theta_{n - 1} - 2u_n
\theta_{n + 1} ) \nonumber \\ & & + \tilde bk_n (u_n \theta_{n -
1} - 2u_{n + 1} \theta_{n + 1} )  \\
F_u^B (k_n ) & = & (au_{n - 1} + bu_n )k_n u_{n - 1} u_n \label{FuB}\\
F_\theta^B(k_n ) &= &(\tilde au_{n - 1} + \tilde bu_n )k_n
\theta_{n -1} \theta_n \ \label{FTB} 
\end{eqnarray}
where $a$, $b$, $\tilde a$ and $\tilde b$ are positive parameters.
In the SabraT model, $u_n$ and $\theta_n$ are complex variables with~\cite{GuoThesis}:
\begin{eqnarray}
I_u^S(k_n)&=& ik_n \lambda (u^ *  _{n + 1} u_{n + 2} - {\delta
\over 2}u^ *  _{n - 1} u_{n + 1} \nonumber \\ & & + {{1 - \delta }
\over4}u_{n -1} u_{n - 2} ),\\
I_\theta^{S}(k_n)&=&  ik_n (\alpha _1 u^ *  _{n + 1} \theta_{n +
2} + \alpha _2 u_{n + 2} \theta^ * _{n + 1} \nonumber \\ & & +
\beta _1 u^
* _{n - 1} \theta_{n + 1} - \beta _2 u_{n + 1} \theta ^ *  _{n - 1}
\nonumber\\ && - \gamma _1 u_{n - 1} \theta_{n - 2}  -
\gamma _2 u_{n - 2} \theta_{n - 1})  \\
F_u^S(k_n )&= & \lambda {\mathop{\rm Im}\nolimits}
[k_{n - 1} u^ *_{n - 1} u^ *_n u_{n + 1}
\nonumber \\ &&  + (1 - \delta )k_{n - 2} u^ *_{n-2}u^*_{n-1}u_n ] \label{FuS}\\
F_\theta^S (k_n ) &=& {\mathop{\rm Im}\nolimits} [\gamma _1
(k_{n } u_{n - 1} \theta_{n - 2} \theta^*_n  + k_{n+1} u_n
\theta_{n - 1} \theta^*_{n + 1} )  \nonumber \\ &&- \beta _2k_n
u^*_{n + 1} \theta_{n - 1} \theta_n  + \gamma _2 k_{n } u_{n - 2}
\theta_{n - 1} \theta^*_n ] \label{FTS}
\end{eqnarray}
The parameters $\alpha_{1,2}$, $\beta_{1,2}$ and $\gamma_{1,2}$ are
determined by
\begin{eqnarray} 
\nonumber \alpha _1 = 4\tau \ , \ \ \beta _1 &=& {1 - \delta }  - 2\tau  \ ,  \ \
\gamma _1 =  - \tau,  \\ 
\alpha _2 = 2 - 4\tau \ ,\ \ \beta _2 &=& 1 - 2\tau \ , \ \
\gamma _2  = \tau - \frac{1 - \delta }{2}  \ \
\label{Eq22}  
\end{eqnarray}
with three free parameters $\lambda$, $\delta$ and $\tau$. In
particular, we fix $\lambda=2$ and $\tau=0.7$ and vary
$\delta$. The value $\delta=1$ is the boundary value separating two
families of Sabra model: a family of three-dimensional-like models
for $0<\delta<1$ and a family of two-dimensional-like models
for $1< \delta<2$. We focus on $0<\delta<1$ in this paper.

We study the scaling behavior of the velocity and temperature structure functions,
$\langle |u_n|^p \rangle$ and $\langle |\theta_n|^p \rangle$, with scaling exponents
$\zeta_p$ and $\xi_p$ defined by:
\begin{equation}
\langle |u_n|^p \rangle \sim k_n^{-\zeta_p} \ ; \qquad
\langle |\theta_n|^p \rangle \sim k_n^{-\xi_p}
\label{zetaxi}
\end{equation}
where $\langle \ldots \rangle$ denotes a time average.
The K41 scaling would be characterized by $\zeta_p = \xi_p = p/3$ while
the BO scaling by $\zeta_p = 3p/5$ and $\xi_p = p/5$. 
In our numerical calculations, we integrate 
the equations of motion using fourth order Runge Kutta method 
with an initial condition of $u_n=\theta_n=0$ except for a small perturbation of $\theta_n$ 
at intermediate values of $n$. 
The Brandenburg model is forced with $f_n = f \delta_{n,0}$ where $f$ is a uniform random noise
while the SabraT model is forced with a Gaussian time-correlated noise acting on $n=3$ and $4$
only~\cite{Sabra}. For the results presented in this work, we summarize the parameters 
used in Table~\ref{tab1}.

\begin{table}
\caption{\label{tab1} Values of the parameters used for the
results presented.}
\begin{tabular}{|c|c|c|c|c|c|c|}\hline
\multicolumn{7}{|c|} {Brandenburg model}  \\\hline
  $a$& $b$& $\tilde {a}$ and $\tilde{b}$ &
$\nu$ & $\kappa$ &$\alpha g $&$N$   \\\hline 0.01 &1 &1&$5\times
10^{-17}$&$5\times 10^{-15}$&1&32\\\hline 0.31 &0.6 &1&$5\times
10^{-9}$&$5\times 10^{-9}$&1&25\\ \hline\hline
\multicolumn{7}{|c|} {SabraT model}  \\\hline
 $\delta$ &$\lambda $&$\tau$ &$\nu$ & $\kappa$
&$\alpha g $&$N$\\\hline 0.5 &2 &0.7&$  10^{-8}$&$10^{-8}$&1&23\\
\hline 0.8 &2 &0.7&$  10^{-8}$&$10^{-8}$&1&23\\ \hline
\end{tabular}
\end{table}

\begin{figure}[bth]
\centerline{
\includegraphics[height=4.8cm,angle=0]{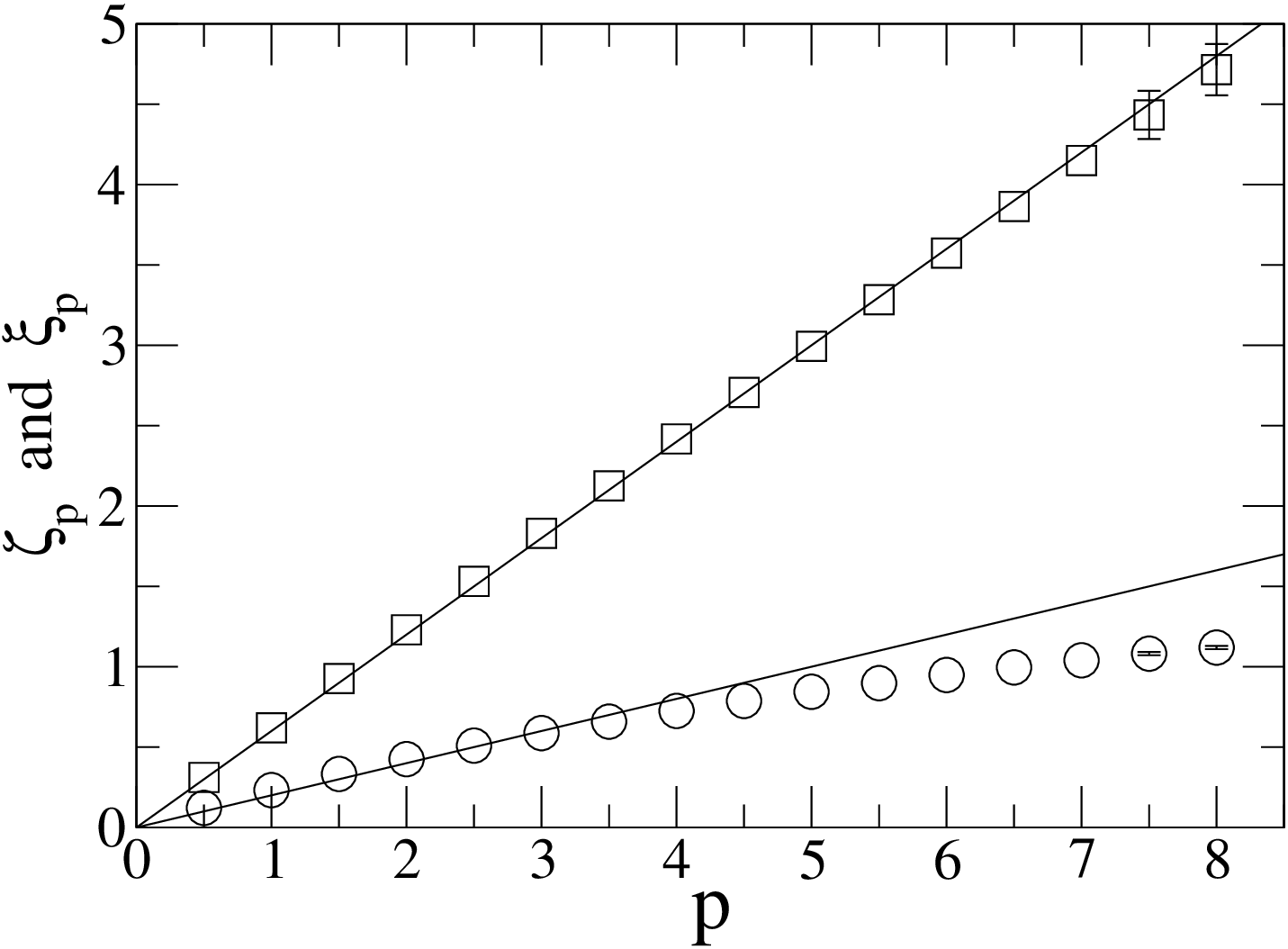}}
\caption{\label{fig1} The scaling exponents $\zeta_p$~(squares) and $\xi_p$~(circles)
for Brandenburg model with $a=0.01$ and $b=1$.
The error increases with $p$ and the largest errors are shown.
Comparing with the two solid lines of slopes $1/5$ and $3/5$ shown, it can be seen that
the scaling behavior is BO with corrections.}
\end{figure}

\begin{figure}[bth]
\centerline{
\includegraphics[height=4.8cm,angle=0]{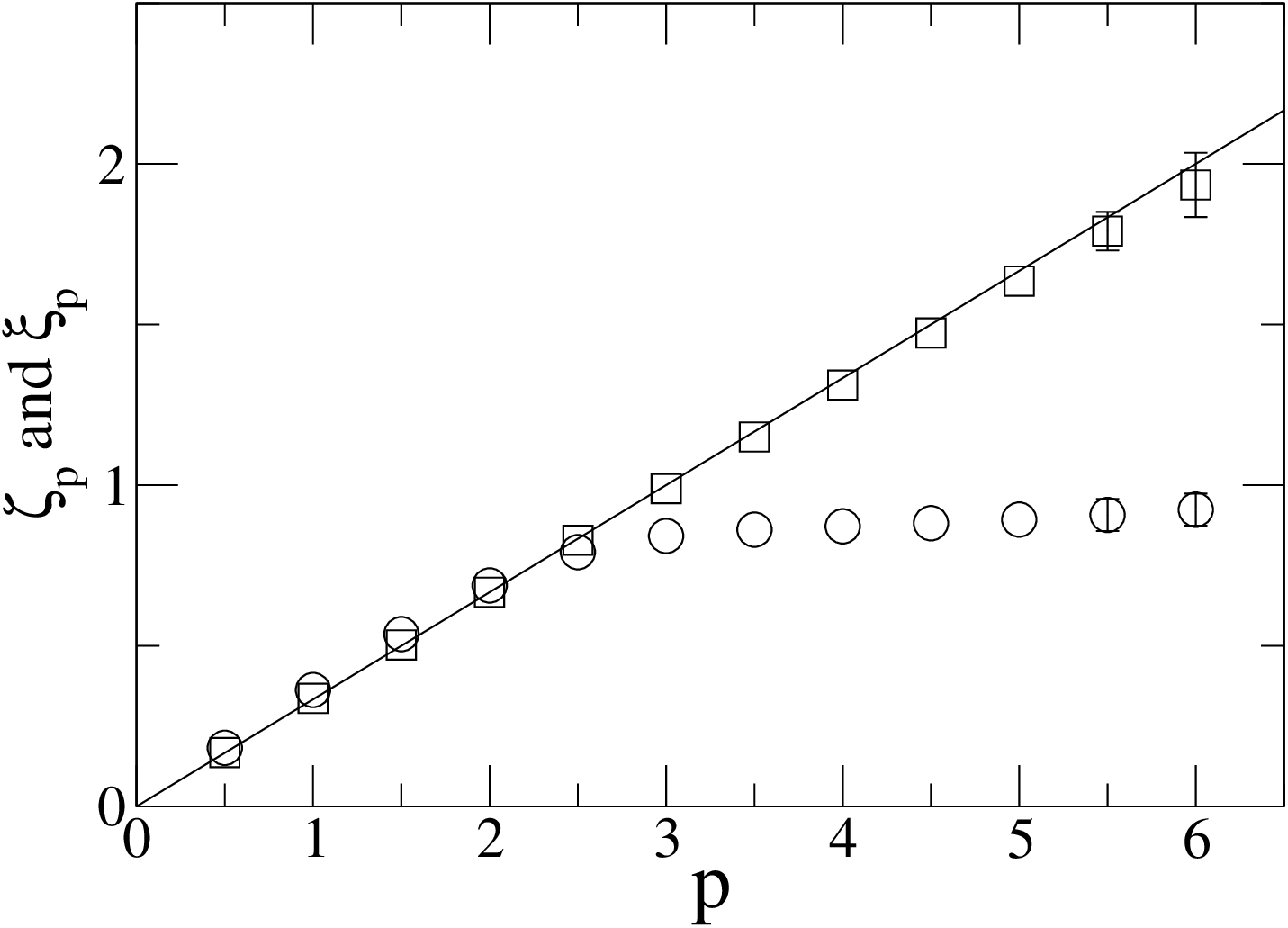}}
\caption{\label{fig2} Same as Fig.~\ref{fig1}
for $a=0.31$ and $b=0.6$. The solid line shown has slope $1/3$.}
\end{figure}

In the Brandenburg model, the scaling behavior depends on the relative magnitudes
of the parameters $a$ and $b$, as reported in earlier studies~\cite{Brandenburg}.
When $b/a$ is larger than some critical value of about 2, 
the scaling exponents $\zeta_p$ and $\xi_p$ are given by
the BO values plus corrections.
The scaling behavior improves with $b/a$.
On the other hand, when $b/a$ is 
smaller but close to the critical value, the scaling exponents
$\zeta_p$ and $\xi_p$ are the same as those obtained 
in the case of passive scalar advection in which the
coupling term $\alpha g \theta_n$ with temperature in the velocity
equation of motion is replaced by a random forcing at $n=0$.
This indicates that buoyancy does not play a
part in the statistical properties in this case.
The scaling exponents for $b/a = 100$ and $b/a=1.94$
are shown respectively in Figs.~\ref{fig1} and \ref{fig2}.
For even smaller values of $b/a$, further away from the critical value,
the system is not chaotic, and in most of the shells the solution is given instantaneously by
the fixed-point solution of $u_n = A k_n^{-1/3}$ and $\theta_n = B k_n^{-1/3}$, which 
holds exactly in the limit of large $N$ and $\nu =\kappa = \alpha g = 0$. 

\vspace{0.5cm}

\vspace{0.5cm}

\begin{figure}[bth]
\includegraphics[height=4.8cm,angle=0]{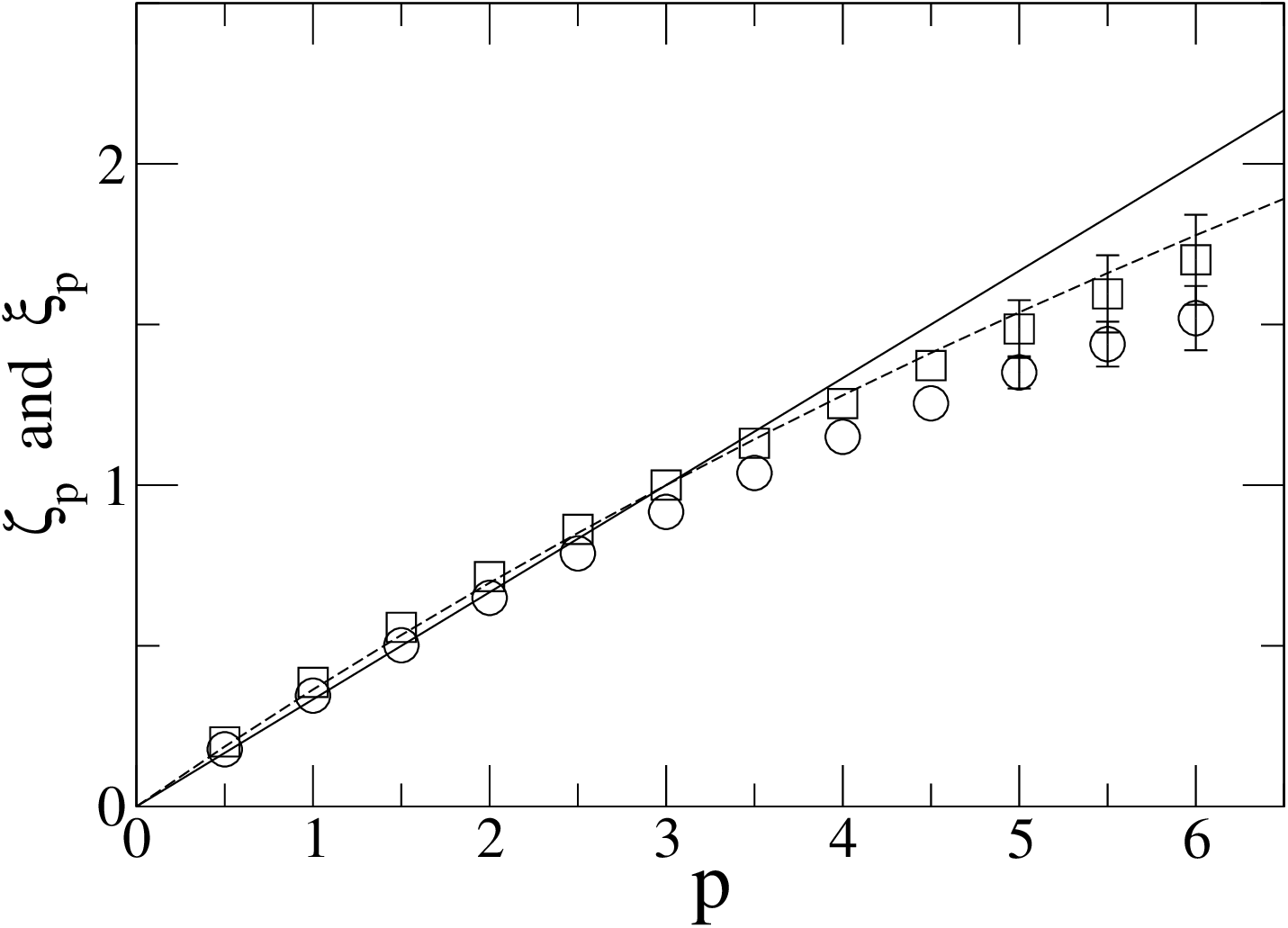}
\caption{\label{fig3} Same as Fig.~\ref{fig2} for the SabraT model with $\delta=0.5$.
The solid line shown has a slope of $1/3$ while the dashed line is the She-Leveque result\cite{SL}.}
\end{figure}

For the SabraT model, we find that the values of $\zeta_p$ 
remain the same as those in the Sabra model without the coupling term $\alpha g \theta_n$
for all the values of $\delta$ studied, again indicating that buoyancy does not play a role
in determining the statistical properties in the SabraT model for $0 < \delta < 1$. 
The precise values of $\zeta_p$ depend on $\delta$, as was reported in 
the GOY model~\cite{GOYKLWB}. In Fig.~\ref{fig3}, we present the results for
$\zeta_p$ and $\xi_p$ for $\delta = 0.5$, a conventional value at 
which the model conserves helicity in the inviscid
limit~\cite{GOYKLWB}. In this case, the values of $\zeta_p$ are 
well described by the She-Leveque result~\cite{SL} of $\zeta_p = p/9 + 2[1-(2/3)^{p/3}]$, 
as was also reported~\cite{Liu} for the GOYT model with $\delta = 0.5$.

\vspace{0.5cm}

\section{\label{scale} The buoyancy scale}

In this section, we discuss how to determine whether buoyancy is
relevant or not in determining the statistical properties.
Consider Eq.~(\ref{Eu}), which is the energy budget.
The third term of the right hand side
is the rate of energy dissipation in the $n$th shell due to viscosity
while the last term is
the power injected into the $n$th shell by the buoyancy forces.
It is thus reasonable to
take buoyancy to be significant in the $n$th shell if
\begin{equation}
|\alpha g \langle Re\{u_n \theta_n^*\} \rangle| > \epsilon
\label{buoyant}
\end{equation}
where $\epsilon \equiv \nu \sum_n k_n^2 \langle |u_n|^2 \rangle$ is the average energy
dissipation rate. We denote the scale at which the equality sign in Eq.~(\ref{buoyant})
holds to be the buoyancy scale $k_{n^*}$. Hence buoyancy is relevant and
significant for $n < n^*$ and irrelevant or insignificant for $n > n^*$.
It is easy to show that for $u_n$ and $\theta_n$
satisfying exactly K41 or BO scaling, $k_{n^*} = 1/L_B$,
where $L_B \equiv \epsilon^{5/4} \chi^{-3/4} (\alpha g)^{-3/2}$
is the Bolgiano length~\cite{Monin} and
$\chi$ is the average thermal or entropy dissipation rate
given by $\chi \equiv \kappa \sum_n k_n^2 \langle |\theta_n|^2 \rangle$.

\begin{figure}[bth]
\centerline{
\includegraphics[height=4.8cm,angle=0]{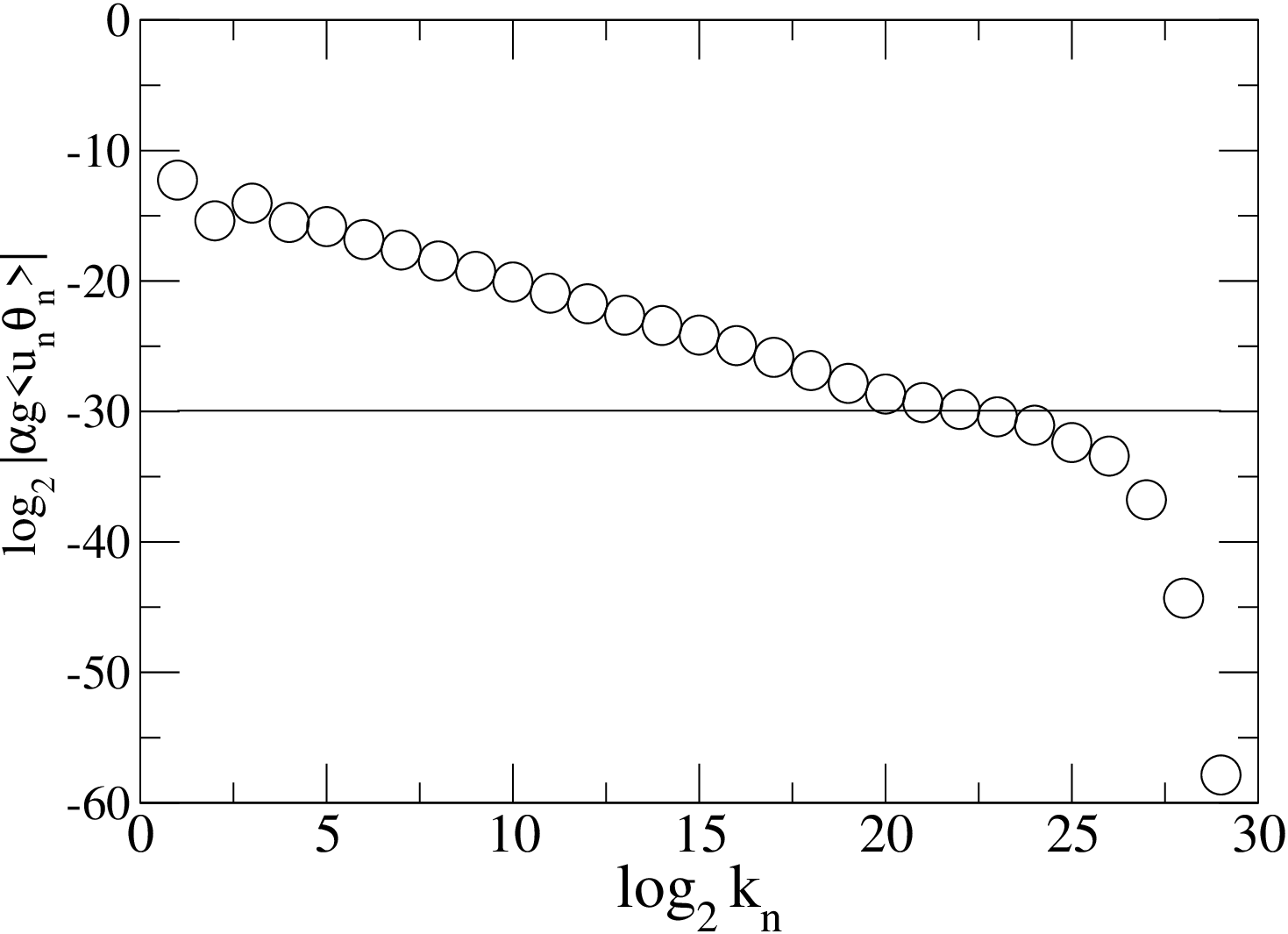}}
\caption{\label{fig4} Comparison of $|\alpha g \langle
{u_n\theta_n} \rangle|$~(circles) with $\epsilon$~(solid line) in
each shell for the Brandenburg model with a large value of $b/a=100$.}
\end{figure}

\begin{figure}[bth]
\centerline{
\includegraphics[height=4.8cm,angle=0]{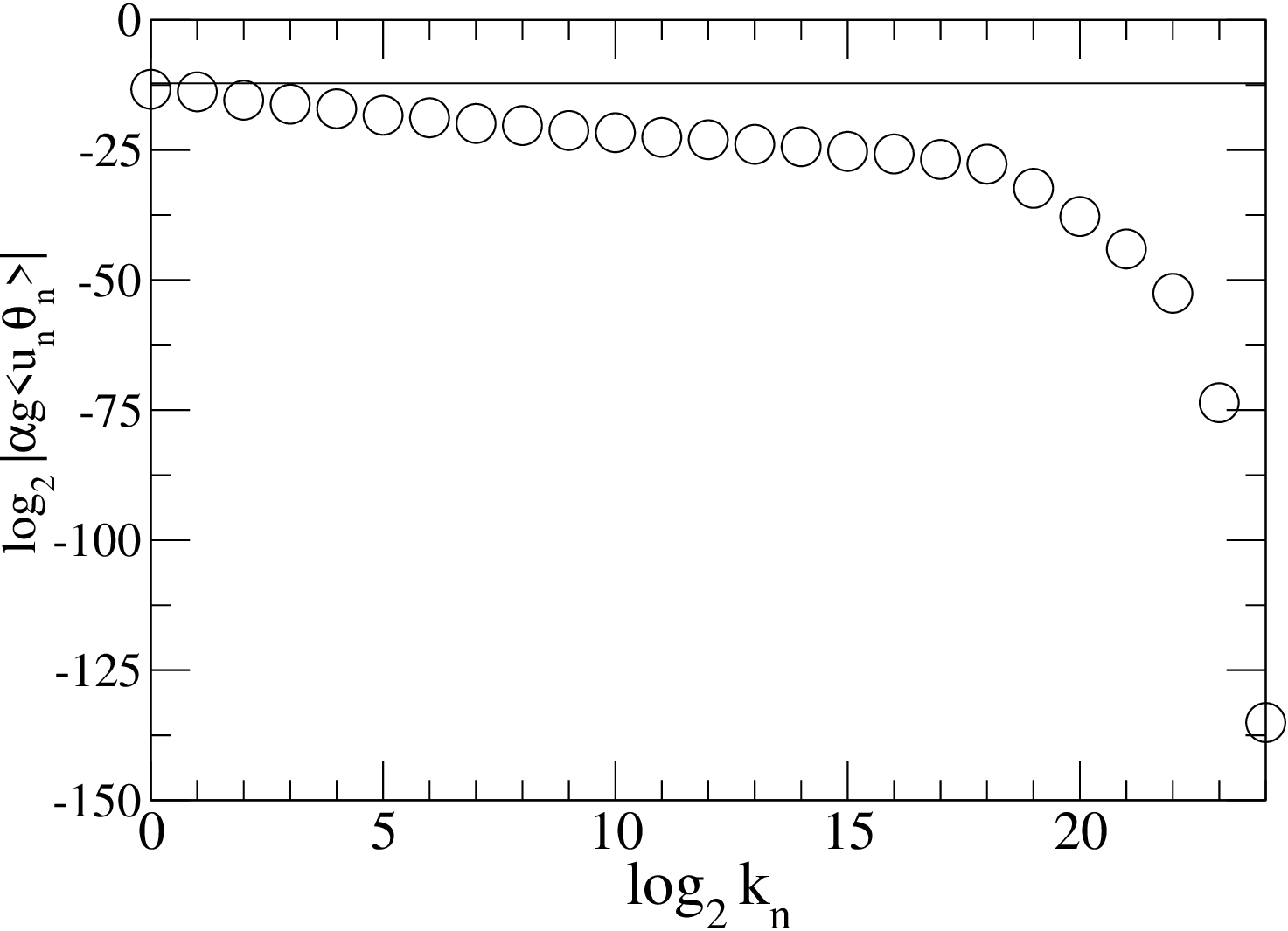}}
\caption{\label{fig5}
Same as Fig.~\ref{fig4} for a small value of $b/a \approx 1.9$.}
\end{figure}

\begin{figure}[bth]
\includegraphics[height=4.8cm,angle=0]{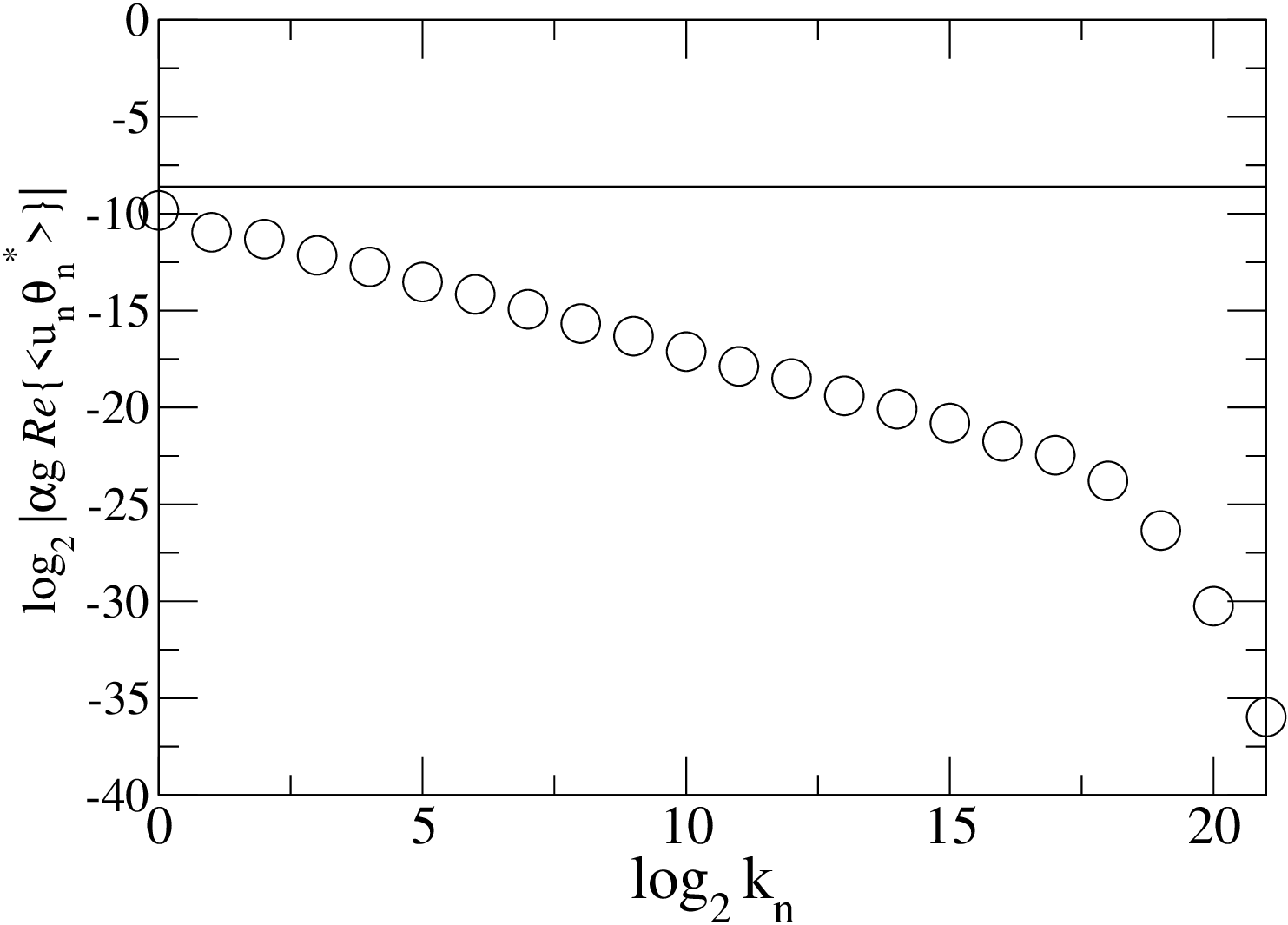}

\vspace{0.8cm}

\includegraphics[height=4.8cm,angle=0]{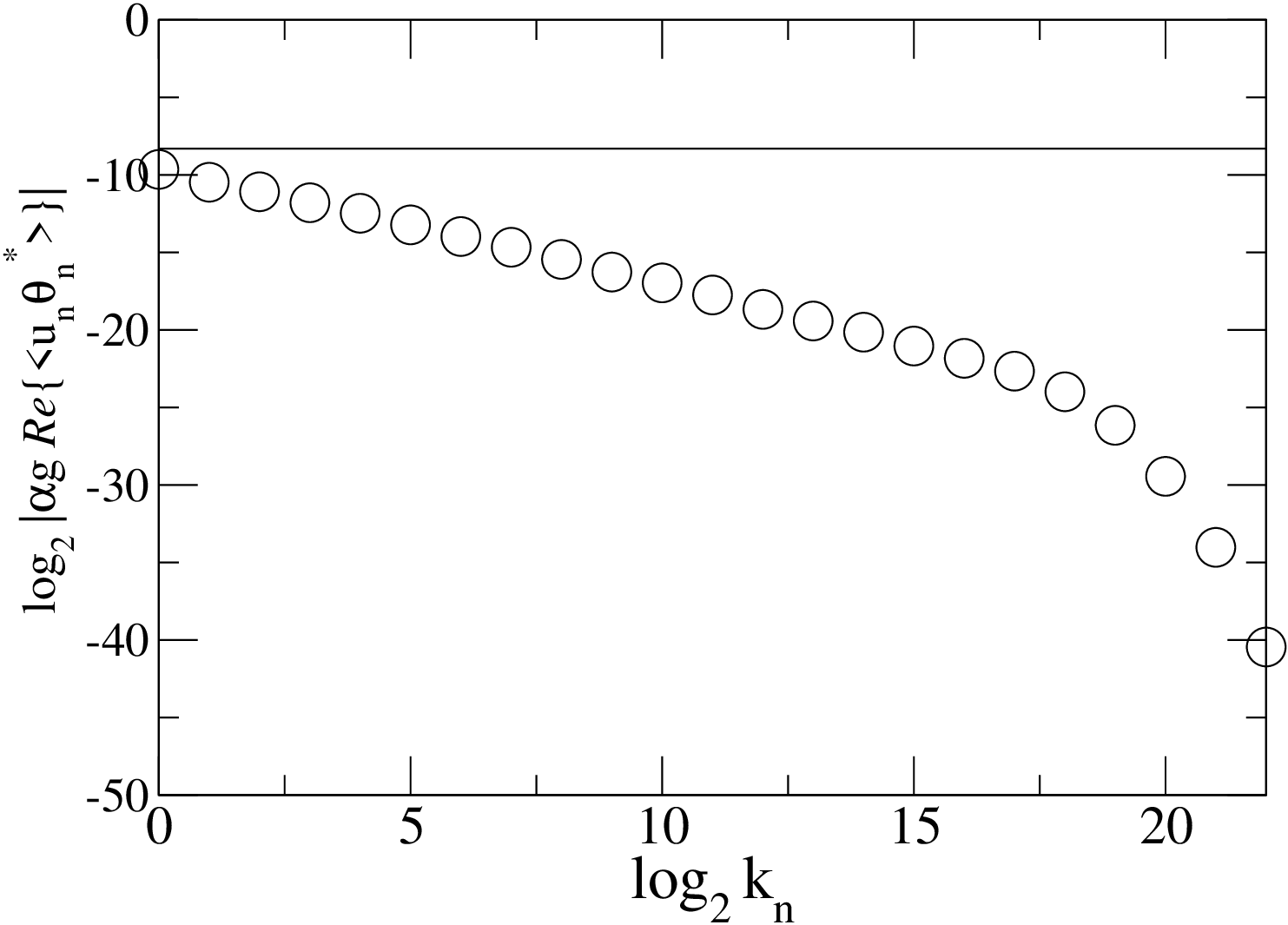}
\caption{\label{fig6}
Comparison of  
$|\alpha g \langle Re\{u_n \theta_n^*\} \rangle|$~(circles)
with $\epsilon$~(solid line) in
each shell for the SabraT model with 
$\delta=0.5$ in the top panel 
and $\delta=0.8$ in the bottom panel.}
\end{figure}

As shown in Figs.~\ref{fig4} and \ref{fig5}, 
we find that 
Eq.~(\ref{buoyant}) is satisfied for most of the shells only in
the Brandenburg model with $b/a$ larger than the critical value.
When $b/a$ is smaller than the critical value,
buoyancy is insignificant in all except the largest shells.
For the SabraT model, we find that buoyancy is insignificant in all except the largest shells
for all the values of $\delta$ studied. The results for $\delta=0.5$ and $\delta=0.8$ are shown in
Fig.~\ref{fig6}.

\vspace{0.5cm}

\vspace{1.5cm}

One naturally expects
different scaling behavior when buoyancy is significant
and when it is not. In this sense, it is not puzzling that
different scaling behavior was reported in the the various shell models proposed.
Indeed we find BO scaling plus corrections when buoyancy is significant and
K41 scaling plus correction when it is not.
The two different scaling behavior
can be understood by studying the evolution equations of energy
and entropy. In the intermediate range where external forcing is not acting and where
energy and entropy dissipation rates are both small,
Eqs.~(\ref{Eu}) and (\ref{ET}) can be approximately written as:
\begin{eqnarray}
\label{Eu2} F_u(k_n)- F_u(k_{n+1})+
\alpha g Re\{ u_n \theta^* _n \} &\approx&0  \\
\label{ET2} F_\theta(k_n)- F_\theta(k_{n+1})&\approx& 0 
\end{eqnarray}
From Eq.~(\ref{ET2}), $F_\theta(k_n)$ is independent of $k_n$ in
the intermediate range, implying that there is an entropy cascade.
From Eq.~(\ref{Eu2}), we see that $\alpha g Re \{u_n \theta_n^*
\}$ is comparable with $F_u$ when buoyancy is significant, and
$F_u(k_n) - F_u(k_{n+1}) \approx 0$ when buoyancy is
insignificant. Thus when buoyancy is insignificant, there is 
also an energy cascade as in the usual inertia-driven turbulence.

\vspace{0.5cm}

In the case when buoyancy is significant, there is only the cascade of entropy.
As a result, one expects the statistical properties to be
controlled by the entropy cascade.
Specifically, one expects~\cite{ChingCheng} the statistical 
properties of $u_n$ and $\theta_n$ 
to be determined solely by $F_\theta$, $\alpha g$, and $k_n$:
\begin{eqnarray}
|u_n| &=& \phi_u (\alpha g)^{2/5} |F_\theta(k_n)|^{1/5} k_n^{-3/5}
\label{RSH1u}\\
|\theta_n| &=& \phi_\theta (\alpha g)^{-1/5} |F_\theta(k_n)|^{2/5} k_n^{-1/5}
\label{RSH1theta}
\end{eqnarray}
where $\phi_u$ and $\phi_\theta$ are dimensionless random variables that are independent of
$k_n$ and statistically independent of $F_\theta(k_n)$. 
On the other hand, when buoyancy is insignificant, there is also the
energy cascade. Thus one expects the statistical properties of $u_n$ and $\theta_n$ 
to be determined by $F_u$, $F_\theta$ and $k_n$:
\begin{eqnarray}
|u_n| &=& \psi_u |F_u(k_n)|^{1/3} k_n^{-1/3}
\label{RSH2u} \\
|\theta_n| &=& \psi_\theta |F_u(k_n)|^{-1/6}|F_\theta(k_n)|^{1/2} k_n^{-1/3}
\label{RSH2theta}
\end{eqnarray}
where $\psi_u$ and $\psi_\theta$ are dimensionless random variables that are independent of
$k_n$ and statistically independent of $F_u(k_n)$ and $F_\theta(k_n)$.
Hence we have
\begin{eqnarray}
\langle |u_n|^p \rangle &\sim& \langle |F_\theta(k_n)|^{p/5} \rangle k_n^{-3p/5} 
\label{uBO}\\
\langle |\theta_n|^p \rangle &\sim& \langle |F_\theta(k_n)|^{2p/5} \rangle k_n^{-p/5} 
\label{TBO}
\end{eqnarray}
when buoyancy is significant and
\begin{eqnarray}
\langle |u_n|^p \rangle &\sim& \langle |F_u(k_n)|^{p/3} \rangle k_n^{-p/3} 
\label{uK41}\\ 
\langle |\theta_n|^p \rangle &\sim& \langle |F_u(k_n)|^{-p/6} |F_\theta(k_n)|^{p/2} \rangle 
k_n^{-p/3} 
\label{TK41}
\end{eqnarray}
when it is not. Equations (\ref{uBO}) and (\ref{TBO}), and Eqs. (\ref{uK41}) and (\ref{TK41})
thus respectively give BO and K41 scaling plus intermittency corrections for the case when
buoyancy is significant and when it is not, just as what was found numerically. 
Moreover, when buoyancy is significant, 
the intermittency corrections are solely due to fluctuations in $F_\theta$ while in the
case when buoyancy is insignificant, the intermittency corrections are due to fluctuations 
in both and $F_u$ and $F_\theta$. We have checked and verified~\cite{ChingCheng}
Eqs.~(\ref{uBO}) and (\ref{TBO}).

\vspace{0.5cm}

\begin{figure}[h]
\centerline{
\includegraphics[height=5.0cm,angle=0]{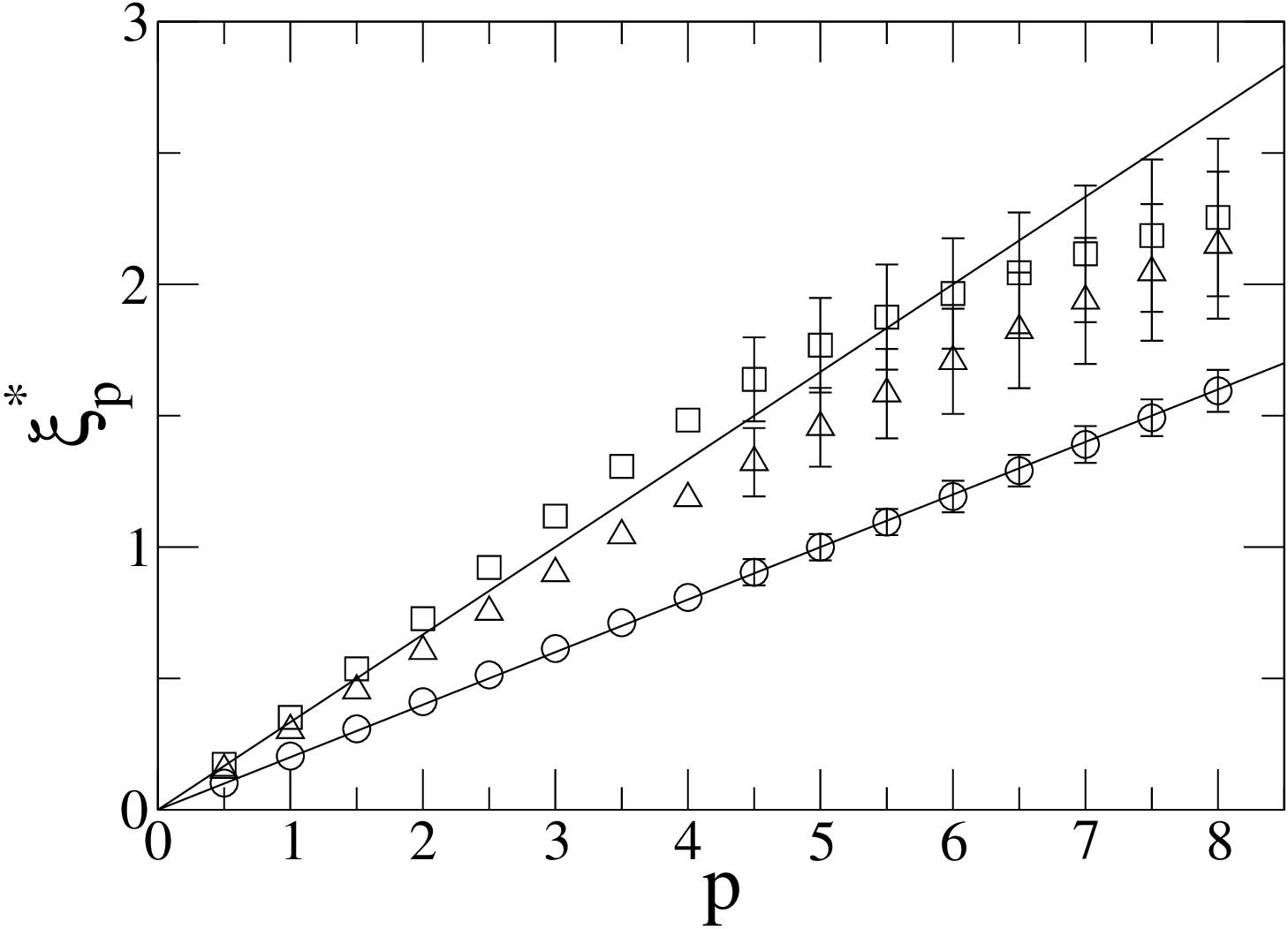}}
\caption{\label{fig7} The scaling exponents $\xi_p^*$ of
the conditional temperature structure functions
$\langle |\theta_n|^p \ \big| \ F_\theta = x \rangle$ for
the Brandenburg model with
small $b/a$~(squares) and large $b/a$~(circles), and the SabraT model with $\delta=0.5$~(triangles).
The error increases with $p$ and the largest errors are shown.
Two solid lines with slopes $1/3$ and $1/5$ are shown.}
\end{figure}

Our work shows that the mere presence of a coupling term
between velocity and temperature in the equations of motion does
not automatically imply that buoyancy is significant and affects
the statistical properties. This leads to the question: How can one tell
whether temperature is behaving as an active or a passive scalar
in models for turbulent thermal convection?
For shell models, one can use Eq.~(\ref{buoyant}).
If Eq.~(\ref{buoyant}) is satisfied in most shells then buoyancy is significant and
temperature is active otherwise temperature would behave as a passive scalar.
It would also be useful to have some other criterion that involves 
directly the statistical features of temperature.
Equations (\ref{RSH1u}) and (\ref{RSH1theta}) imply that when buoyancy is significant, 
the conditional statistics of $u_n$ and $\theta_n$ at fixed values of $F_\theta$
would have simple BO scaling with no corrections~\cite{ChingCheng}.
On the other hand, this is not true when buoyancy is insignificant; instead
Eqs.~(\ref{RSH2u}) and (\ref{RSH2theta}) indicate that
the conditional statistics of $u_n$ and $\theta_n$ at fixed values of $F_\theta$
continue to deviate from simple K41 scaling. 
Hence one can study the conditional statistics of temperature 
at fixed values of the entropy transfer rate. 
If these conditional statistics are described by simple scaling 
then temperature is behaving as an active scalar. 
Otherwise if the conditional statistics remain anomalous then temperature 
is behaving as a passive scalar.  To check this idea,
we calculate the the conditional temperature structure functions
at fixed values of entropy transfer rate and their scaling exponents
$\xi_p^*$:
\begin{equation}
\langle |\theta_n|^p \ \big| \ F_\theta = x \rangle \sim k_n^{-\xi_p^*}
\end{equation}
for the SabraT model the Brandenburg model for both small and large values of $b/a$.
The results of $\xi_p^*$ do not depend on $x$ and are shown in Fig.~\ref{fig7}.
It can be seen that for Brandenburg model with large $b/a$, $\xi_p^*$ are
indeed well described by the BO values of $p/5$. Also, as expected, 
for both the Brandenburg model with small
$b/a$ and the SabraT model, $\xi_p^*$'s continue to deviate from the K41 values of $p/3$.

\section{\label{final}Conclusions}

Various shell models have been proposed for turbulent thermal
convection. K41 scaling plus corrections has been reported in most
of these models while BO scaling plus intermittency corrections is reported in
the Brandenburg model with suitable parameters. In this paper, we
have shown that buoyancy is not always significant and relevant to the statistical properties 
in these shell models even though there is an explicit coupling term
with temperature in the equation of motion for velocity. We have
further clarified that 
BO scaling plus corrections would be observed 
only in the shell models in which
buoyancy is significant. For shell models in which buoyancy is insignificant, the statistical properties
remain the same as in the case in which the coupling term with temperature is absent.
We have argued that the statistics properties
are controlled solely by the cascade of entropy when buoyancy is significant 
but controlled by both the cascades of energy and entropy when buoyancy is not significant,
and shown how this leads to the two different scaling behavior in the two cases.
We have further shown that the intermittency corrections are solely attributed to fluctuations of
the entropy transfer rate when buoyancy is significant but are caused 
by fluctuations of both the energy and entropy transfer rate when buoyancy is insignificant. 
As a result, the conditional temperature structure
functions at fixed entropy transfer rate would have simple scaling when buoyancy is significant but
remain anomalous when buoyancy is insignificant. We have demonstrated how this feature 
can be used as a criterion to distinguish whether temperature is acting as an active or a passive scalar.




\begin{acknowledgments}
This work is supported by the Hong Kong Research Grants Council~(CUHK
400304 and CA05/06.SC01).
\end{acknowledgments}


\end{document}